# Suppression of Jahn-Teller distortion by chromium and magnesium doping in spinel $LiMn_2O_4$: A first-principles study using GGA and GGA+U

*Gurpreet Singh[1], S. L. Gupta[2], R. Prasad[2*], S. Auluck[2], Rajeev Gupta[2] and Anjan Sil[1]*

[1.] *Department of Materials and Metallurgical Engineering, Indian Institute of Technology, Roorkee, Uttarakhand, India-247667.*

[2.] *Department of Physics, Indian Institute of Technology, Kanpur, Uttar Pradesh, India-208016.*

## Abstract

The effect of doping spinel $LiMn_2O_4$ with chromium and magnesium has been studied using the first-principles spin density functional theory within GGA (generalized gradient approximation) and GGA+U. We find that GGA and GGA+U give different ground states for pristine $LiMn_2O_4$ and same ground state for doped systems. For $LiMn_2O_4$ the body centered tetragonal phase was found to be the ground state structure using GGA and face centered orthorhombic using GGA+U, while for $LiM_{0.5}Mn_{1.5}O_4$ (M= Cr or Mg) it was base centered monoclinic and for $LiMMnO_4$ (M= Cr or Mg) it was body centered orthorhombic in both GGA and GGA+U. We find that GGA predicts the pristine $LiMn_2O_4$ to be metallic while GGA+U predicts it to be the insulating which is in accordance with the experimental observations. For doped spinels, GGA predicts the ground state to be half metallic while GGA+U predicts it to be insulating or metallic depending on the doping concentration. GGA+U predicts insulator-metal-insulator transition as a function of doping in case of Cr and in case of Mg the ground state is found to go from insulating to a half metallic state as a function of doping. Analysis of the charge density and the density of states suggest a charge transfer from the dopants to the neighboring oxygen atoms and manganese atoms. We have calculated the Jahn-Teller active mode displacement $Q_3$ for doped compounds using GGA and GGA+U. The bond lengths calculated from GGA+U are found to be in better agreement with the experimental bond lengths. Based on the bond lengths of metal and oxygen, we have also estimated the average oxidation states of the dopants.

## I. Introduction

Spinel $LiMn_2O_4$ and layered $LiMnO_2$ are being used as lithium ion battery cathode materials [1-6]. $LiMn_2O_4$ is known to be cubic spinel at room temperature. As one lowers the temperature the cubic spinel transforms to orthorhombic at $280^0K$ [7]. As this transition temperature is very close to the room temperature, continuous re-cycling results in a rapid fall in the capacity of the battery and hence a poor performance of the battery. On further lowering the temperature to $65^0K$ the structure totally transforms to a single phase of tetragonal spinel. It is therefore very important to stabilize the cubic spinel structure. Various divalent and trivalent ions, which make the structure more stable, have been tried as dopants by different groups [8-10]. Experimentally, among the various possible dopants chromium and magnesium have been studied extensively [9, 10]. Chromium doped spinels have shown to work successfully in the higher voltage range. On the other hand magnesium was found to be successful in suppressing the oxygen non-stoichiometry in $LiMn_2O_4$, which was a necessary condition for the structural transition of $LiMn_2O_4$ near room temperature [11]. A large amount of quantitative computational work on lithium manganese oxides (primarily $LiMnO_2$) have been carried out by Prasad *et al*. [12 - 16] and Mishra and Ceder [17]. However, not much computational work has been done on the effect of dopants on spinel $LiMn_2O_4$. A study on chromium doped $LiMn_2O_4$ using the local density approximation (LDA) has been carried out by Shi *et al*. [18]. Spin polarized (anti-ferromagnetic) generalized gradient approximation (GGA) calculations were performed only for the end compounds such as $LiMn_2O_4$ and $LiCrMnO_4$. Mishra and Ceder, in their study on structural stability of lithium manganese oxides, have stressed the use of GGA. In the LDA the Jahn-Teller distortions are meta-stable or unstable [17]. Hence in the present study all the calculations are based on spin polarized GGA and GGA+U. Shi *et al*. performed calculations on un-relaxed systems. It was concluded that there is a very slight increase in the charge density around manganese atoms even for the maximum doping content of chromium. A major change in the charge density was found for the oxygen atoms. It was also observed that the shape of the density of states (DOS) for both manganese and oxygen atoms remains almost unchanged after doping with chromium. It would, therefore, be interesting to see if these conclusions are valid after the unit cell parameters and the atomic positions are relaxed. In the present study we report calculations on the relaxed structures. In fact we find that there is an increase in the charge density around some of the manganese atoms with increase in chromium doping. The shape of the DOS changes with the concentration and nature of the dopant.

To the best of our knowledge we are not aware of any theoretical work on the effect of magnesium doping in $LiMn_2O_4$. Only few experimental studies on magnesium as dopant are available. It has been observed that Jahn-Teller distortion and charge ordering contribute towards the structural transition of $LiMn_2O_4$ at low temperatures [8-10, 19]. In the present work an attempt has been made to study the effect of dopants on the structural stability and electronic properties of pristine $LiMn_2O_4$ through first-principles methods. This is motivated by the experimental work carried by Singh *et al*. [20, 21] on chromium and magnesium doping. Chromium is anti-ferromagnetic and magnesium is non-magnetic. Chromium has an atomic radius comparable to manganese. So, there is a chance that chromium easily replaces manganese experimentally. Magnesium is a lighter element and is, therefore, of great interest from lithium battery point of view. Electronic properties and structural changes in pristine $LiMn_2O_4$ and doped with chromium and magnesium have been studied using the density functional theory (DFT).

The plan of the paper is as follows. In Section II, we give the computational details. In Section III, our results are presented and discussed for both pristine $LiMn_2O_4$ and doped systems. Finally in Section IV we give our conclusions.

## II. Computational details

The structure optimization has been done within the DFT [22, 23] using VASP (*Vienna ab initio simulation package)* [24-27] and the projector augmented wave method (PAW) pseudo-potentials [28]. The Kohn-Sham equations [29, 30] are solved using the exchange correlation function of Perdew and Wang [31] for generalized gradient approximation (GGA) and GGA+U. We chose the Hubbard parameter U to be 4.5 eV and 5.0 eV for chromium and manganese respectively guided by Wang *et al.* [32]. The exchange energy is fixed at 1 eV. We have used the automatic Monkhorst-Pack scheme [33] for the generation of k-points (8x8x8). For optimizing the structure, a conjugate gradient algorithm (CGA) [34] is used. The tetrahedron method with Blöchl corrections [35] is used for k-integrations to determine the total energy. All the calculations are done at $0^oK$. The plane wave cutoff was set at 550 eV to ensure the good convergence of stress tensor during cell-parameter relaxation. All the structures were fully relaxed.

The conventional unit cell of $LiMn_2O_4$ has 56 atoms. To speed up our calculations we have used the primitive unit cell with 2 formula units i.e. 14 atoms. The primitive unit cell is chosen using the lattice vectors as $\vec{a}' = -\vec{a}+\vec{b}+\vec{c}$, $\vec{b}' = \vec{a}-\vec{b}+\vec{c}$ and $\vec{c}' = \vec{a}+\vec{b}-\vec{c}$. This transformation modifies the fractional co-ordinates. This transformation results from the fact that the arrangement of ions remains invariant in the Cartesian co-ordinate system as given by,

$$x\vec{a} + y\vec{b} + z\vec{c} = x'\vec{a}' + y'\vec{b}' + z'\vec{c}'$$

The optimized relaxed structures were obtained by repeated sequential relaxing of the volume, the ion positions and the shape of the primitive unit cell.

The primitive cell for $LiMn_2O_4$ contains two lithium, four manganese and eight oxygen atoms. Lithium atoms occupy the 8a tetrahedral sites (0.125, 0.125, 0.125) and manganese atoms occupy the 16d octahedral sites (0.5, 0.5, 0.5) while the oxygen atoms form a cage and occupy the 32e sites (x, x, x). We can distinguish the four manganese atoms by their positions:

Mn(I) (0.5, 0.5, 0.5); Mn(II) (0.5, 0.0, 0.5); Mn(III) (0.0, 0.5, 0.5); Mn(IV) (0.5, 0.5, 0.0)

The starting lattice parameters and the oxygen position parameters were taken from an earlier experimental study [36]. It was found that Mn (I) and Mn(IV) lie in the (110) plane while Mn(II) and Mn(III) lie in the (220) plane.

We have done the calculations on the ferromagnetic and anti-ferromagnetic ordering of various manganese atoms to ascertain the ground state. GGA calculations showed ferromagnetic ordering to be more stable than anti-ferromagnetic ordering. The energy difference between them is 0.26eV. This is not in agreement with the experimental studies. Further, each manganese atom interacts with its nearest manganese via $90^o$ interactions intermediated by oxygen atoms. The Goodenough – Kanamori rules [37] suggest that the interaction should be anti-ferromagnetic (AFM). Ouyang *et al.* [38] have shown that GGA+U is necessary to obtain the AFM ground state. We have, therefore, performed spin polarized calculations for the AFM ordering of the manganese atoms. It was found that the total energy of the system remains same in all the AFM configurations irrespective of the direction of spin assigned to the manganese atoms for pristine $LiMn_2O_4$. So, out of the four manganese atoms we can assign any two manganese atoms as spin-up and the other two as spin-down.

In the doped compounds chromium and magnesium were placed at the octahedral sites, which were earlier occupied by manganese atoms. The site preference of the dopants was checked by placing them at various possible sites. It was observed that total energy of the system remains unchanged in the various configurations. The minimum doping concentration for a 14 atoms cell is x = 0.5. The compositions were taken to be $LiM_xMn_{2-x}O_4$ (x=0.5, 1.0 and M= Cr, Mg). Experimentally, it has been shown that doping is beneficial for lithium ion battery up to a limited extent of dopants [9]. So we have also restricted ourselves up to x=1.0. Since $LiCr_2O_4$ is a hypothetical oxide, no experimental data is available for this compound even though some theoretical calculations have been reported in the literature [39]. We have performed calculations for the following compositions:

(i) $LiMn_2O_4$; (ii) $LiCr_{0.5}Mn_{1.5}O_4$; (iii) $LiCrMnO_4$; (iv) $LiMg_{0.5}Mn_{1.5}O_4$; (v) $LiMgMnO_4$

## III. Results and Discussion

### A. Pristine $LiMn_2O_4$

We first consider the pristine $LiMn_2O_4$ system in cubic spinel structure with the primitive unit cell containing 14 atoms and experimental lattice constant 8.241 Å. This structure was optimized using the procedure discussed in Section II, keeping the shape fixed. This gave a lattice constant of 8.218 Å and 8.376 Å using GGA and GGA+U respectively. It has been found experimentally that $LiMn_2O_4$ changes to orthorhombic phase near 285°K [40-42]. We have performed GGA calculations on orthorhombic phase and GGA+U calculations both on orthorhombic and tetragonal phase. For simplicity, we have used the structural information given in Ref. 42 for the orthorhombic phase. Our results are in agreement with the earlier studies done by Grechnev *et al.* [44] and Ouyang *et al.* [38].

The final optimized structures obtained after relaxing the shape for GGA and GGA+U are different. The final optimized structure for GGA is body centered tetragonal ($I4_1/amd$) with a= 5.851 Å and c = 8.123 Å and for GGA+U is face centered orthorhombic with a = 8.332 Å, b = 8.382 Å and c = 8.416 Å. Though, the lattice constants of body centered tetragonal and orthorhombic primitive unit cell seem to be different, their volumes are nearly same. Lattice parameter values in the latter case show that the b/a and c/a ratios are close to 1. The optimized structural parameters for all the systems studied are given in Table I. These results are found to be consistent with the earlier studies done on the system [17].

The total DOS for pristine $LiMn_2O_4$ in cubic, orthorhombic and tetragonal structure are shown in Fig. 1. We see that GGA predicts the ground state to be metallic, whereas GGA+U predicts the ground state to be insulating. We note that the GGA+U prediction is in agreement with the experimentally observed behaviour. For both GGA and GGA+U, the DOS is dominated by the manganese atoms in all the three phases. Thus the GGA+U drives the system away from the metallic behavior.

In Fig. 2 we have plotted the $t_{2g}$ ($d_{xy}$, $d_{yz}$ and $d_{xz}$) and $e_g$ ($d_{z^2}$ and $d_{x^2-y^2}$) components of the partial DOS for optimized pristine $LiMn_2O_4$ from (i) GGA and (ii) GGA+U. DOS for Mn(I) and Mn(II); Mn(III) and Mn(IV) is found to be similar so we have shown the DOS only for Mn(I) and Mn(III) in both cases. The partial dos for $t_{2g}$ spin-up states moves towards the higher energy side while the $t_{2g}$ spin-down moves to lower energy side in GGA+U compared to GGA (Fig. 2). This is consistent with the fact the inclusion of U introduces onsite repulsion.

## B. Doped LiMn$_2$O$_4$

In the pristine structure the Mn atoms were replaced by M (Cr and Mg) one by one so as to obtain LiCr$_x$Mn$_{2-x}$O$_4$ and LiMg$_x$Mn$_{2-x}$O$_4$ for x=0.5, 1.0. All the structures were optimized after doping. For GGA and GGA+U, the final optimized structure for LiM$_{0.5}$Mn$_{1.5}$O$_4$ was found to be base centered monoclinic and for LiMMnO$_4$ it was found to be body centered orthorhombic. Fig. 3 shows the total DOS of the doped systems for (i) GGA and (ii) GGA+U respectively. GGA predicts that Cr doping brings the system towards a semimetallic state via a half metallic state (Fig. 3(i) a, b), where as, GGA+U gives a metallic state for x=0.5 and an insulating state for x=1.0 for Cr (Fig. 3(ii) a, b). This is interesting that GGA+U predicts an insulator-metal-insulator transition as a function of Cr doping. The partial DOS for the manganese atoms in LiCr$_x$Mn$_{2-x}$O$_4$ (x=0.5, 1.0) obtained from GGA and GGA+U are shown in Figs. 4 ((i) a-d, e-h) and Figs. 5 (a-d, e-h). The splitting of $t_{2g}$ orbitals is clearly observed in both cases. Inclusion of U has moved the spin-up and spin-down states away from the Fermi level and results in the insulating state for GGA+U and a semi metallic state for GGA. Both GGA and GGA+U predict that Mg doping leads the system towards a half metallic state (Figs 3(i) c, d) and Figs. 3((ii) c, d) respectively. On doping with magnesium (x=0.5) the density of states around the Fermi level is found to be less compared to LiMn$_2$O$_4$ for the GGA calculations. Finite spin-up density of states at the Fermi level is present, but the Fermi level lies well in the gap for the spin-down states for x=1.0.

Both GGA and GGA+U predict that in case of chromium doped $LiMn_2O_4$ (for x=0.5), the number of electrons in the $e_g$ state of Mn(III) are less compared to the pristine $LiMn_2O_4$ and the number of electrons in the $t_{2g}$ state of Mn(III) are more compared to the pristine system. This shows that there is a transfer of electrons from $e_g$ to $t_{2g}$ states on Cr doping. We see from Fig. 4 that as the chromium content is increased (x=1.0), GGA calculations shows that the number of electrons present in the lower energy states of Mn(III) are higher than for x=0.5. These results are in agreement with the charge density values obtained from the VASP output, which shows the charge density to be larger by 0.016 electronic charges for Mn(III) in $LiCrMnO_4$ than $LiCr_{0.5}Mn_{1.5}O_4$. However, in case of GGA+U, both spin-up and spin-down density of states are found to be absent around the Fermi level as the chromium content increases from x= 0.5 to 1.0. This results in insulator-metal-insulator transition.

The GGA results for the partial DOS plots of magnesium doped for x=0.5 (Fig. 5(i)), show a clear splitting of the $t_{2g}$ states into triplet for the case of Mn(II) atom. The $t_{2g}$ (down) states of Mn(II) were found to be totally filled. Even though a clear splitting of $e_g$ can be seen from the plots, these states were found to be empty. A similar trend was also found in other manganese atoms. As the content of magnesium is increased to x= 1.0, the splitting of the $t_{2g}$ states almost disappears and the number of electrons in the lower energy states decreases. The energy gap between the $t_{2g}$ (up) and $t_{2g}$ (down) states also decreases (Fig. 5). From GGA+U calculations it can be seen that in case of x=0.5, the Fermi level shifts towards the lower energy value and it touches the top of the valence band. In case of x=1.0, this shifts even more towards the lower energy and it lies well within the $t_{2g}$ energy states. The energy gap between the $t_{2g}$ (up) and $t_{2g}$ (down) states is found to increase with the increase in magnesium content.

Charge density plots in the (110) plane for the various compositions are shown in Fig. 6 for x=0.5. Both GGA and GGA+U predict the transfer of charge from the transition metal to the surrounding oxygen atoms. This continues to be valid for higher doping concentrations and is consistent with DOS. Thus we have shown the results for GGA only.

We now discuss the suppression of Jahn-Teller (JT) distortion by Cr and Mg doping. In $LiMn_2O_4$ the JT effect occurs because manganese exists in $Mn^{3+}$ and $Mn^{4+}$ states. $Mn^{3+}$ is a JT active ion; therefore there will be a distortion of the octahedron in this case. Since Cr and Mg are not JT ions, their substitutional doping is expected to suppress JT distortion. This suppression is further enhanced because some $Mn^{3+}$ ions get promoted to $Mn^{4+}$ ions which are not JT ions. We can analyze the JT distortion in terms of Jahn-Teller amplitudes (discussed below).

Van Vleck [45] in his study on JT effect showed that there exist three symmetrical modes of vibration of an octahedral complex $XY_6$ as $\nu_1$ ($Q_1$), $\nu_2$ ($Q_2,Q_3$) and $\nu_3$ ($Q_4,Q_5,Q_6$). The magnitude of elongation is proportional to the amplitude $Q_3$ of $e_g$ (JT active) frozen mode. Values of $Q_1$, $Q_2$ and $Q_3$ are given as follows

$Q_1 = [u_x(1)-u_x(4)+u_y(2)-u_y(5)+u_z(3)-u_z(6)]/\sqrt{6}$

$Q_2 = [u_x(1)-u_x(4)-u_y(2)+u_y(5)]/2$

$Q_3 = [2u_z(6)-2u_z(3)-u_x(1)+u_x(4)-u_y(2)+u_y(5)]/\sqrt{12}$

where the value of $u_i(j)$ refers to atomic displacement (bond length changes) with respect to the perfect undistorted octahedron. Values of $Q_3$ (JT active) frozen mode were plotted (Fig. 7) against breathing mode $Q_1$ for each manganese atom in each composition. The circle is in the figure is a guide to the eyes for the effective JT mode for various compositions. For GGA, it can be seen from the $Q_1$ and $Q_3$ values that for manganese atoms in chromium doped $LiMn_2O_4$ are smaller compared to the magnesium doped $LiMn_2O_4$. GGA+U calculations show that the $Q_1$ values for the manganese atoms in magnesium doped $LiMn_2O_4$ are less compared to GGA. Thus, GGA predicts Cr to be more effective than Mg in suppressing the JT distortion while GGA+U predicts Mg to be more effective.

The Mn-O bond length in spinel $LiMn_2O_4$ (un-relaxed structure) was found to be same for all Mn-O bonds in the octahedron (1.945 Å). In case of GGA after relaxing the structures two different kinds of bond lengths were obtained in $LiMn_2O_4$. One with $d_1^{Mn-O}$ =1.965 Å and other with $d_2^{Mn-O}$ =1.921 Å. These results are consistent with the earlier experiments which also reported two different bond lengths [47]. In case of GGA+U calculations after relaxing the structure two different kinds of bond lengths were obtained in $LiMn_2O_4$. One with $d_1^{Mn-O}$ = 1.932 Å and other with $d_2^{Mn-O}$ = 2.021 Å. These results are in good agreement with the study carried by Carvajal *et al.* [43]. Our results differ from the recent work of Ouyang *et al.* who found that the ground state of pristine $LiMn_2O_4$ to be orthorhombic for GGA and GGA+U while we find it to be body centered tetragonal using GGA and orthorhombic using GGA+U. An effort was made to determine the oxidation state of doped metal ion by comparing the average bond lengths of M-O (M = Mg, Cr, Mn) obtained from first principles study using VASP with predictions based on tabulated ionic radii. The bond length of the M-O was calculated by assuming the oxygen ionic radii to be 1.40 Å and various possible hypothetical oxidation states of each metal ion. Manganese exists in two possible states, $Mn^{3+}$ and $Mn^{4+}$. $Mn^{3+}$ can be further subdivided into two possibilities of low spin and high spin. In low spin configuration all the four electrons will be present in the $t_{2g}$ states, but in the high spin configuration one of the electrons among the four jumps into $e_g$ state. This causes the JT distortion in the structure. Chromium can also have two possible valence states 3+ and 4+ but it is not a JT active ion. Magnesium can exist in 2+ valence state only. A graph between the average M-O bond lengths obtained from VASP and calculated for different oxidation states [48] is shown in Fig. 8 for GGA and GGA+U. The solid line represents the exact match between the calculated values and those obtained from the first principles calculations. It is clear from the figure that manganese and chromium are possibly present in 3+ and 4+ oxidation states as the points corresponding to these are close to the solid line.

Two different bond lengths predicted by GGA are close to the values of theoretical bond lengths of $Mn^{3+}$ low spin (L.S.) and $Mn^{4+}$ respectively. On doping chromium (x=0.5) four kinds of Mn-O bond lengths were found *viz.* 1.92, 1.94, 1.95 and 1.97 Å. It has been observed that the number of Mn-O bonds with bond length 1.97 Å was decreased. This shows that manganese is moving away from 3+ oxidation state. On increasing chromium content it has been observed that there exist only two kinds of Mn-O bond lengths *viz.* 1.94 Å and 1.92 Å. The theoretical bond length of $Mn^{4+}$ - O is 1.93 Å. This shows that chromium has shifted the oxidation state of manganese from +3 to +4. In case of magnesium doping (x=0.5) there exist two kinds of Mn-O bond lengths *viz.* 1.91 Å and 1.96 Å. On increasing the magnesium content the Mn-O bond lengths are observed to be 1.82 and 1.93 Å.

Two different bond lengths predicted by GGA+U are close to the values of theoretical bond lengths of $Mn^{3+}$ high spin (H.S.) and $Mn^{4+}$. $Mn^{3+}$ (H.S) is a Jahn-Teller active ion, which causes the distortion of the spinel structure. Therefore, it can be concluded that GGA+U is giving a clearer picture of the Jahn-Teller distortion in the cubic spinel structure. On doping chromium (x=0.5) four kinds of Mn-O bond lengths were found *viz.* 1.93 and 1.95 Å; 1.96 and 1.97 Å. On further increasing the content of chromium it has been found that only two kinds of bond lengths remain in the structure *viz.* 1.93 Å and 1.95 Å. This shows that the chromium is driving the manganese oxidation states towards $Mn^{+3}$ (L.S.) and $Mn^{4+}$ and hence suppressing the Jahn-Teller distortion. On doping magnesium in $LiMn_2O_4$, it has been observed that there exist two kinds of Mn-O bond lengths *viz.* 1.93 Å and 1.97 Å. This shows that manganese is present in $Mn^{+4}$ and $Mn^{+3}$ (L.S.). As the content of magnesium increases the Mn-O bond lengths are found to be 1.91 Å and 1.94 Å.

## V. Conclusions

We have studied the electronic structure of pristine and doped $LiMn_2O_4$ using GGA and GGA+U methods. Our results show that GGA and GGA+U predict different ground states for pristine $LiMn_2O_4$. GGA gives body centered tetragonal phase and GGA+U leads to face centered orthorhombic ground state. Both GGA and GGA+U give the same ground state for doped systems. Base centered monoclinic for $LiM_{0.5}Mn_{1.5}O_4$ (M = Cr, Mg) and body centered orthorhombic for $LiMMnO_4$ (M = Cr, Mg). GGA+U predicts an insulating ground state for pristine $LiMn_2O_4$. Our study based on GGA+U shows the presence of an insulator-metal-insulator transition as a function of chromium doping, whereas the magnesium doping brings the system to a half metallic state. Suppression of splitting of the $e_g$ states for doped systems shows the suppression of JT distortion. GGA+U shows the effectiveness of the magnesium doping in suppressing the JT distortion while GGA shows chromium to be more effective for suppression of JT distortion. Both GGA and GGA+U predicts the significant charge transfer between the manganese atoms and the dopant atoms. An attempt was made to calculate the average oxidation state of manganese, chromium and magnesium present in the various structures. Two types of Mn-O bond lengths were found in the relaxed structure of pristine LiMn$_2$O$_4$ for GGA ($d_1^{Mn-O}$ = 1.21 Å and $d_2^{Mn-O}$ = 1.965 Å) and GGA+U ($d_1^{Mn-O}$ = 1.932 Å and $d_2^{Mn-O}$ = 2.021 Å). The values predicted by GGA+U are in better agreement with the experimental results. This shows the presence of two kinds of manganese ions with different ionic radii. Chromium was found to be present in oxidation states of 3+ while magnesium was found to be present in 2+ oxidation state in the doped systems. Thus the content of Mn$^{3+}$ in doped systems is less compared to the pristine compound which implies less JT distortion in doped system. Thus, the doping of LiMn$_2$O$_4$ with chromium and magnesium increases the structural stability.

Table I: Lattice parameters of the various systems studied.

| Various system studied | LiMn$_2$O$_4$ | | LiCr$_{0.5}$Mn$_{1.5}$O$_4$ | | LiMg$_{0.5}$Mn$_{1.5}$O$_4$ | | LiCrMnO$_4$ | | LiMgMnO$_4$ | |
|---|---|---|---|---|---|---|---|---|---|---|
| | GGA | GGA+U | GGA | GGA+U | GGA | GGA+U | GGA | GGA+U | GGA | GGA+U |
| a(Å) | 5.851 | 8.332 | 10.007 | 10.162 | 9.965 | 10.064 | 5.775 | 5.831 | 5.723 | 5.697 |
| b(Å) | 5.851 | 8.382 | 5.778 | 5.991 | 5.782 | 5.825 | 5.915 | 5.971 | 6.015 | 6.291 |
| c(Å) | 8.123 | 8.416 | 17.274 | 17.517 | 23.761 | 23.987 | 8.156 | 8.320 | 8.369 | 8.368 |
| Cos(β) | 0.000 | 0.000 | -0.831 | -0.829 | -0.846 | -0.847 | 0.000 | 0.000 | 0.000 | 0.000 |

# Figure Captions

FIG. 1: DOS of LiMn$_2$O$_4$ in optimized (a) Cubic, (b) Orthorhombic and (c) Tetragonal phases for (i) GGA and (ii) GGA+U. For orthorhombic structure in GGA, the spin down DOS overlaps the spin up DOS.

FIG. 2: Partial DOS of the manganese atoms in relaxed LiMn$_2$O$_4$. (a) Mn(I) – (0.5, 0.5, 0.5) (b) Mn(III) – (0.0, 0.5, 0.5) for (i) GGA and (ii) GGA+U. Solid line is for t$_{2g}$ (up), dashed line for t$_{2g}$ (down), dots for e$_g$ (up) and dotted line for e$_g$ (down).

FIG. 3: Total DOS of (a) LiCr$_{0.5}$Mn$_{1.5}$O$_4$, (b) LiCrMnO$_4$, (c) LiMg$_{0.5}$Mn$_{1.5}$O$_4$ and (d) LiMgMnO$_4$ for spin up and spin down for (i) GGA and (ii) GGA+U.

FIG. 4: Partial DOS of the chromium [(a) Cr(I) – (0.5, 0.5, 0.5) for x=0.5: (e) Cr(I) – (0.5, 0.5, 0.5), (f) Cr(II) – (0.5, 0.0, 0.5) for x=1.0] and manganese [(b) Mn(II) – (0.5, 0.0, 0.5), (c) Mn(III) – (0.0, 0.5, 0.5), (d) Mn(IV) – (0.5, 0.5, 0.0) for x=0.5: (g) Mn(III) – (0.0, 0.5, 0.5), (h) Mn(IV) – (0.5, 0.5, 0.0) for x=1.0] in LiCr$_x$Mn$_{2-x}$O$_4$ in case of (i) GGA and (ii) GGA+U. Solid line is for t$_{2g}$ (up), dashed line for t$_{2g}$(down), dots for e$_g$(up) and dash-dots for e$_g$(down).

FIG. 5: Partial DOS of the manganese [(a) Mn(II) – (0.5, 0.0, 0.5), (b) Mn(III) – (0.0, 0.5, 0.5) (c) Mn(IV) – (0.5, 0.5, 0.0) for x=0.5: (d) Mn(III) – (0.0, 0.5, 0.5), (e) Mn(IV) – (0.5, 0.5, 0.0) for x=1.0] in LiMg$_x$Mn$_{2-x}$O$_4$ for (i) GGA and (ii) GGA+U. Solid line is for t$_{2g}$ (up), dashed line for t$_{2g}$ (down), dots for e$_g$ (up) and dotted line for e$_g$ (down).

FIG. 6: Charge density for (a) $LiMn_2O_4$, (b) $LiCr_{0.5}Mn_{1.5}O_4$ and (c) $LiMg_{0.5}Mn_{1.5}O_4$ in the (110) planes for GGA. Contour lines are separated by linear increment of $0.1/\text{Å}^3$. Red region corresponds to high charge density and the blue region corresponds to low charge density. GGA+U gives the similar charge density. Thus not shown here.

FIG. 7: Plots showing the Jahn-Teller active frozen phonon mode $Q_3$ as a function of the breathing mode coordinate $Q_1$ in $LiM_xMn_{2-x}O_4$ for GGA and GGA+U. Triangles are for Cr and squares are for Mg. The circle is a guide to the eye showing suppression of JT mode for Cr doping (filled correspond to $x = 1.0$ and empty correspond to $x = 0.5$).

FIG. 8: Plots showing the difference between calculated bond lengths using hypothetical oxidation state of transition metal ion and bond lengths obtained from the first principles study for GGA and GGA+U. Solid line shows the limit of exact match between the two values. L.S. and H.S. correspond to low spin and high spin configuration respectively.

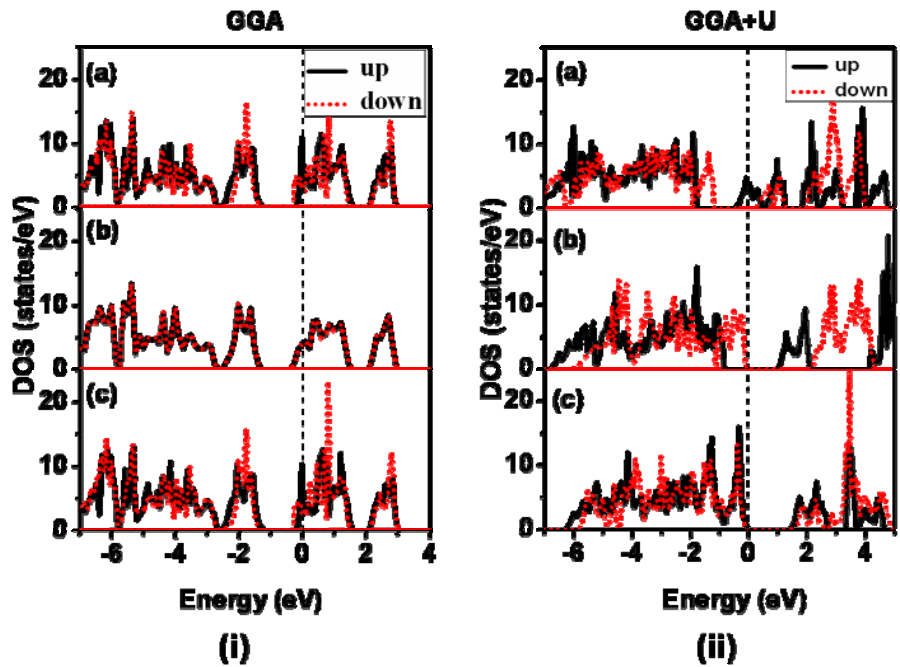

**Figure 1**

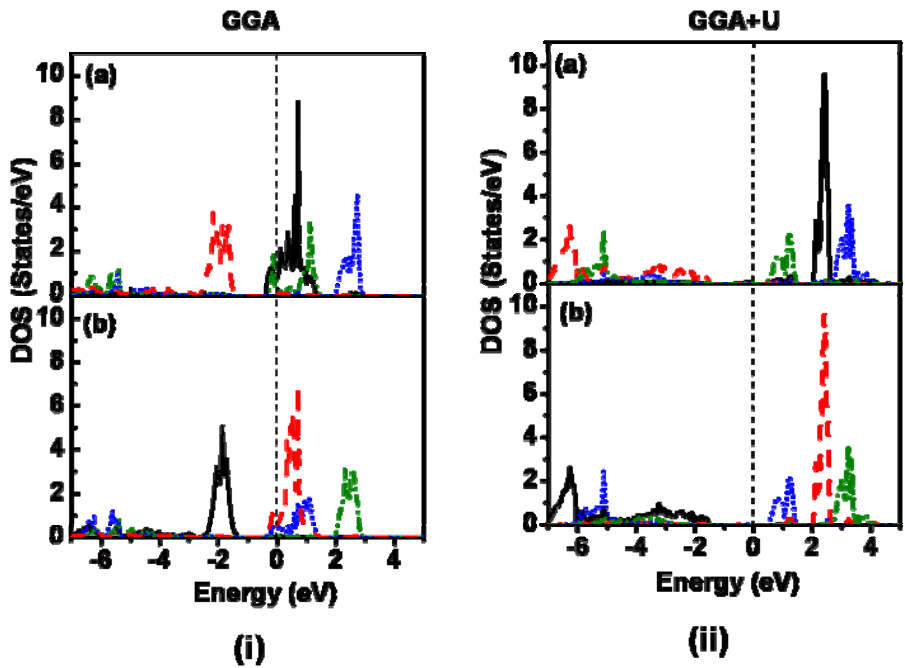

**Figure 2**

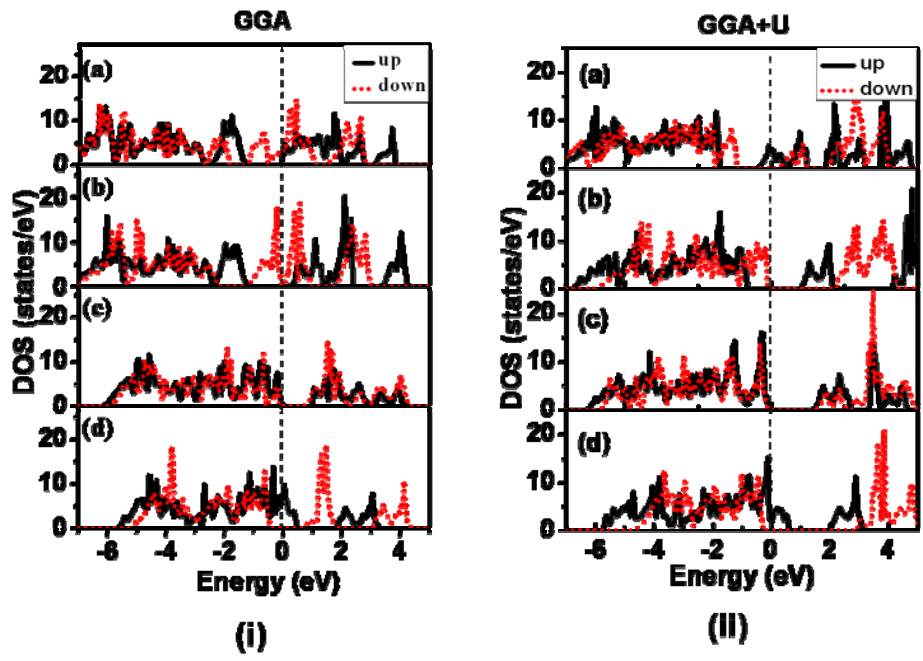

**Figure 3**

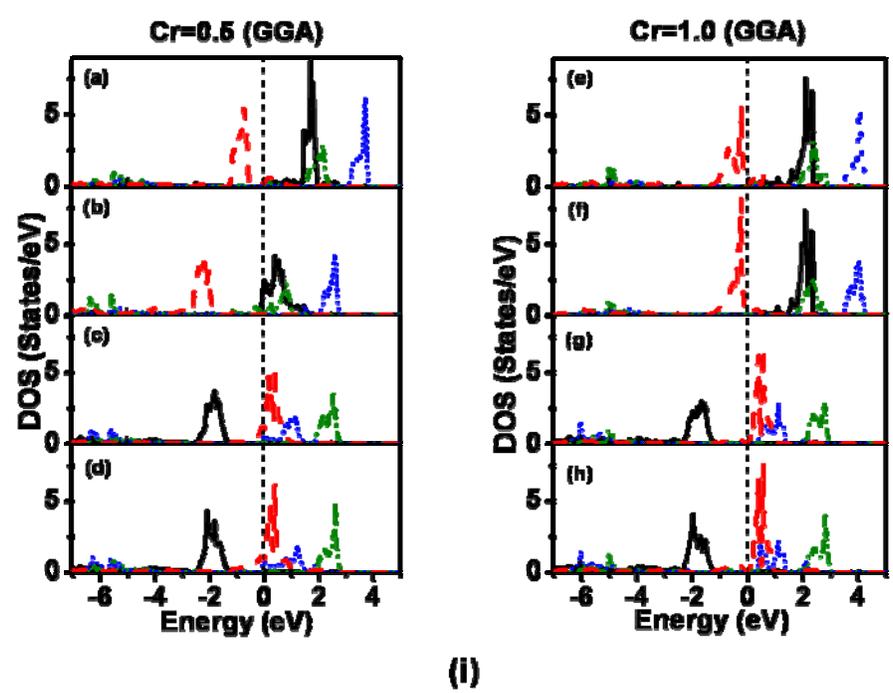

**Figure 4**

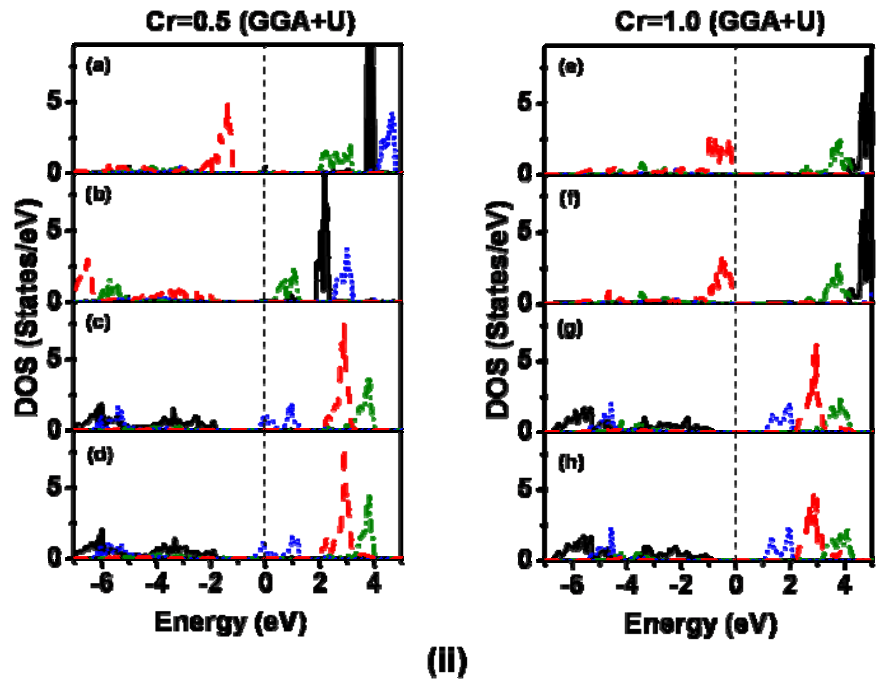

**Figure 4**

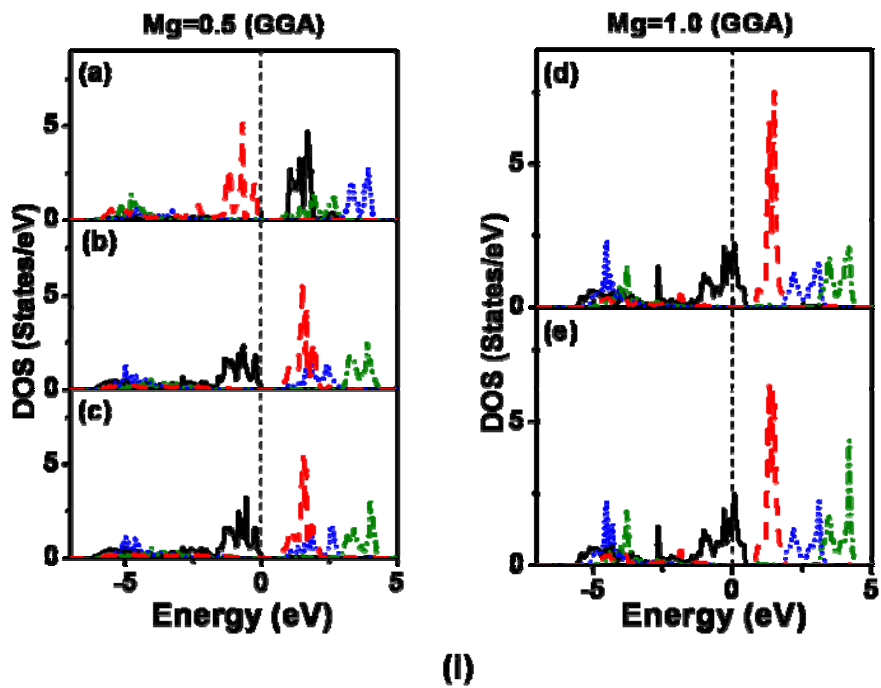

**Figure 5**

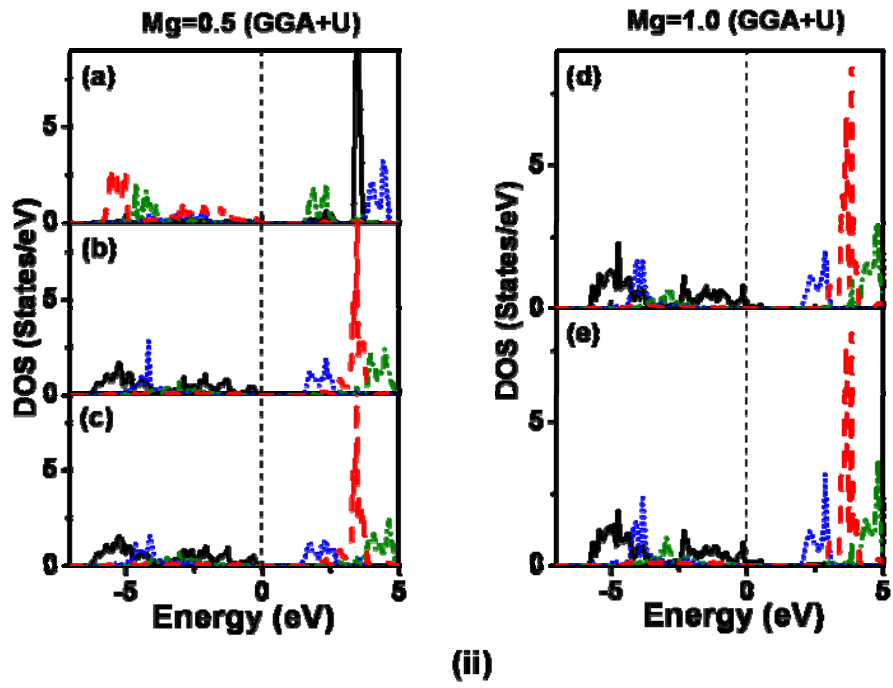

**(ii)**

**Figure 5**

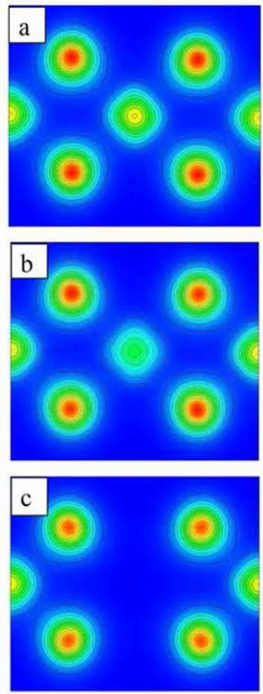

**Figure 6**

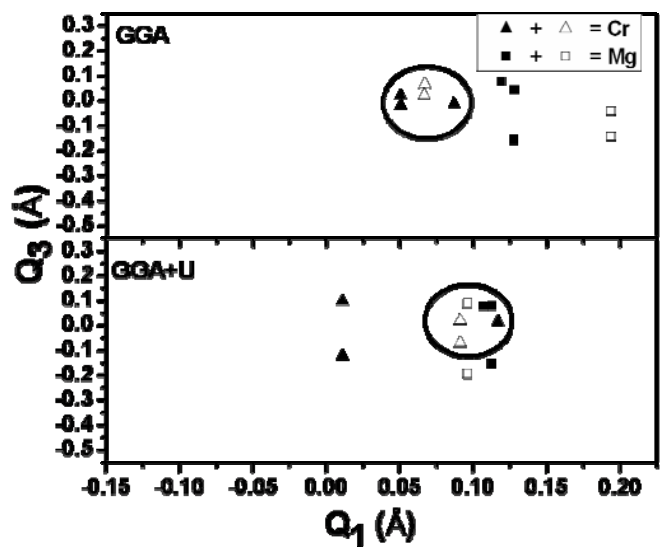

**Figure 7**

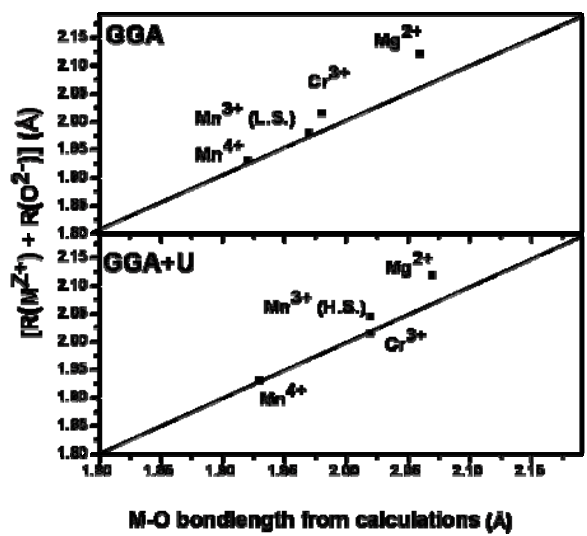

**Figure 8**